\documentclass[sigconf,nonacm]{acmart}

\AtBeginDocument{%
  }

\setcopyright{none}
\renewcommand\footnotetextcopyrightpermission[1]{}
\usepackage{xcolor}

\newif\ifshowchanges
\showchangesfalse

\begin{document}

\title{Lost in Code Generation: Reimagining the Role of Software Models in AI-driven Software Engineering}


\author{Jürgen Cito}
\affiliation{%
  \institution{Software Engineering Group, TU Wien}
  \streetaddress{Favoritenstrasse 9-11}
  \city{Vienna}
  \country{Austria}
  \postcode{1040}
}
\orcid{0000-0001-8619-1271}
\email{juergen.cito@tuwien.ac.at}

\author{Dominik Bork}
\affiliation{%
  \institution{Business Informatics Group, TU Wien}
  \streetaddress{Erzherzog-Johann-Platz 1}
  \city{Vienna}
  \country{Austria}
  \postcode{1040}
}
\orcid{0000-0001-8259-2297}
\email{dominik.bork@tuwien.ac.at}



\renewcommand{\shortauthors}{Jürgen Cito and Dominik Bork}


\begin{abstract}
Generative AI enables rapid ``vibe coding" and agentic software engineering, where natural language prompts yield working software systems. This lowers barriers to software creation, but it also collapses the boundary between prototypes and engineered software. Systems that appear complete may lack robustness, security, and maintainability. We argue that this shift motivates a renewed role for software models. Rather than serving only as upfront blueprints, models can be recovered from AI-generated systems, used to restore comprehension, and refined to guide subsequent evolution. We distinguish an agentic loop, in which recovered models support automated checking and repair, from a human reflective loop, in which models expose assumptions for inspection and revision. We identify candidate model types and illustrate how recovered models can expose implicit assumptions as possible constraints for review and validation. This paper positions software models as a bridge between current software engineering practice and model-based reasoning in AI-driven development.
\end{abstract}





\maketitle

\section{Vibe Coding in Three Acts}

Vibe coding and agentic software engineering are no longer fringe curiosities. Natural-language programming environments, AI coding assistants, and increasingly autonomous development agents are now used to create prototypes, internal tools, production services, and changes to existing codebases. This shift makes the original vibe-coding story less exceptional than it first appeared. 

Alongside this optimism, recent public accounts collect failures across several platforms, including exposed API keys, inverted access control, authentication bypasses, remote code execution, and agentic deletion of production data.\footnote{See the industry overview at \url{https://getautonoma.com/blog/vibe-coding-failures}. We use this as a motivating collection of incidents rather than as a primary empirical study.} The specific cases differ, but they point to the same structural problem. \textit{AI-generated software can be demo-complete and locally plausible while remaining engineering-incomplete, with assumptions about security, data protection, runtime boundaries, and evolution left implicit.}

One early and memorable example was documented by Leonel Acevedo (@leojr94\_ on X), a non-technical creator who built and launched a Software-as-a-Service application using AI coding assistants.\footnote{\url{https://x.com/leojr94_/status/1900767509621674109}. We use this public account as an illustrative anecdote, not as empirical evidence about the prevalence or reliability of vibe coding.}
In a matter of days, he moved from idea to deployed system without writing a single line of code himself. The practice he described as \emph{vibe coding} consisted of expressing functionality in natural language, iteratively pasting instructions into tools such as Cursor,\footnote{\url{https://cursor.com/}}, and accepting the generated outputs as the foundation of his product. Remarkably, this process was not just an academic experiment. Acevedo acquired paying customers and operated a real online business.

\paragraph{\textbf{Act I, The Good}}
This account captures much of the optimism surrounding generative AI in software engineering. It highlights a world where barriers to entry are radically lowered, where anyone can instantiate working software from a loosely expressed idea, and where entrepreneurial imagination can translate into deployed systems without requiring years of technical training. Or, as Matt Welsh in a recent viewpoint provocatively asks, do we face \textit{``The End of Programming''}?~\cite{Welsh23} Vibe coding resonates strongly with decades of aspirations for model-driven engineering, low-code, and citizen development~\cite{brambilla2017model,simon2022low}. Generative AI seems to fulfill parts of these promises in a way that previous paradigms never fully achieved. Instead of painstakingly learning abstractions, users can directly describe intent and obtain running systems~\cite{von2006democratizing, Beheshti24}.

\paragraph{\textbf{Act II, The Bad}}
However, the story does not end with celebration. As Acevedo's user base expanded, familiar difficulties emerged. The application started to suffer from performance bottlenecks, scalability issues, and brittleness under load. What initially appeared as a production-ready system became a prototype in disguise. The system \emph{looked} complete, but its underlying quality attributes had not been engineered. Section~\ref{sec:prototyping} develops this parallel in more detail.

\paragraph{\textbf{Act III, The Ugly}}
The narrative took a further turn when Acevedo began reporting serious security problems. Malicious actors discovered ways to bypass subscription checks, exploit exposed API keys, and inject arbitrary data into the database. The product that had seemed a triumph of accessibility and speed now appeared dangerously fragile. In his own words, \textit{``guys, I'm under attack, ever since I started to share how I built my SaaS using Cursor, random things are happening, maxed out usage on API keys, people bypassing the subscription, creating random stuff on db.''}\footnote{\url{https://x.com/leojr94_/status/1901560276488511759}} These reports illustrate the darker side of generative AI for code generation. Superficially functional systems, once exposed to adversarial environments, can collapse catastrophically.

The point is not that one creator made avoidable mistakes. The broader issue is how generative AI currently mediates software creation. When code is generated directly from natural language, no explicit intermediate artifacts are produced that make the system's structure, requirements,\footnote{The prompting history could be seen as an implicit account of requirements, but is often not visible, and even if it were, reading and comprehending this history is time-consuming and prone to error.} assumptions, or limitations visible. What results is software that may work superficially but is essentially a black box, even to its creator.

\paragraph{\textbf{Beyond the Lone Vibe Coder}}
The early vibe-coding narrative centered on solo non-experts, but the same opacity now matters in professional settings. Junior developers often concentrate on achieving functional correctness, paying limited attention to architecture, maintainability, or non-functional concerns~\cite{esposito2025early,motogna2024uncovering}. Even in large-scale systems, experienced engineers often lack sufficient time, perspective, or appropriate abstractions to holistically reason about generated or scaffolded code~\cite{kuutila2020time,burge2003rationale}. In team settings, the challenge is compounded. One must not only make sense of one's own AI-generated code but also understand the fragments generated by colleagues or agents, without always knowing their prompting intent or conversation history.

Public incidents, industry reports, and early empirical studies point to related risks. In one dramatic case, an autonomous AI agent in Replit deleted a production database despite explicit instructions to freeze code changes, even fabricating fake records and hiding its actions~\cite{lemkin2025replit}. Reports from EclipseSource describe recurring difficulties in integrating AI into established development practice. Teams often lack predefined workflows to incorporate AI agents, generic prompts misalign with legacy or domain-specific constraints, and insufficient training or alignment hampers effective adoption~\cite{eclipsesource2023fail}. Finally, in an interview study of engineers who participated in AI-assisted code review, participants described that LLM suggestions sometimes imposed higher cognitive load due to verbosity, lacked explanatory depth, and were adopted only with skepticism when context was missing~\cite{alami2025human}. Together, these accounts suggest that the opacity introduced by direct generation challenges novices and professionals alike.

\paragraph{\textbf{Scope of the Claim}}
We do not claim that vibe coding has replaced disciplined engineering in large, regulated, or safety-critical systems. Established standards, reviews, architecture work, and coordination practices remain essential in such settings. Our target is the growing middle ground in which AI-generated prototypes become consequential, agents produce changes in existing repositories, and teams inherit generated fragments that must be understood, validated, and hardened.

\paragraph{\textbf{Reimagining the Role of Software Models}}
The good, the bad, and the ugly of vibe coding illustrate both the excitement and the risks of generative AI~\cite{nguyen2025generative, NguyenBZGAF24}. They also motivate the central claim of this paper. Software models should not be treated only as upfront design artifacts. In AI-driven software engineering, they can mediate between human intent, generated code, validation, and long-term evolution. We develop this claim through a prototyping analogy, a concrete model-mediated feedback loop, candidate model types, and research tasks that make the proposal actionable.

\section{From Prototyping to Vibe Coding}
\label{sec:prototyping}
The history of software engineering has repeatedly confronted the tension between speed and rigor~\cite{kuutila2020time, berander2005software}. Prototyping emerged as a practice precisely to address this tension. It allowed developers and stakeholders to explore requirements, validate assumptions, and test user interactions long before a full system was engineered. The core value of prototyping lies in its ability to make ideas tangible at an early stage, thus supporting dialogue and reflection. But the very success of prototypes has always carried risks. Once a prototype acquired a polished interface, it became difficult for stakeholders to resist the temptation to treat it as a finished product~\cite{prototype_critics5, prototype_critics6}. The gap between appearance and substance, between what was demonstrated and what was actually engineered, has been a recurring theme in software practice for decades.

\paragraph{\textbf{From Illusion to Reality}}
This historical experience is highly relevant to vibe coding. Generative AI enables the production of software artifacts that go far beyond the scope of traditional prototypes. They are not sketches, partial simulations, or interactive wireframes. They are executable systems that can be deployed to the cloud, connect to payment providers, and serve real users. Their polish, responsiveness, and functionality make them appear indistinguishable from production-grade systems.

Beneath the surface, familiar risks persist. The generated code often lacks considerations of robustness, efficiency, or security~\cite{OltroggeDSAFRPB18}. Edge cases, concurrency issues, and adversarial interactions are rarely accounted for. The illusion is no longer merely visual but also functional, making the gap between prototype and product harder to recognize.

\paragraph{\textbf{Shifting Knowledge Boundaries}}
However, there is an important difference between earlier prototyping experiences and the current wave of AI-assisted development. Traditional prototypes were created by engineers who generally understood which corners were being cut. They knew that data validation had been deferred, that security concerns had not been addressed, or that performance had not been optimized. This knowledge served as a safeguard, even as the temptation to deploy prototypes prematurely persisted.

In vibe coding, by contrast, many creators are non-experts. They may have entrepreneurial vision or domain expertise, but lack the technical background to understand which aspects of the system are fragile or incomplete. For them, the generated system is a black box. If it runs, it is assumed to be correct. If it looks professional, it is assumed to be robust. The epistemic asymmetry between traditional prototyping and AI-driven prototyping is thus profound. What was once a conscious trade-off made by engineers becomes an unconscious risk borne by non-engineers.

\paragraph{\textbf{Acceleration and Amplification}}
Generative AI does more than reproduce the pitfalls of prototyping. It amplifies them. Deployment pipelines, cloud hosting platforms, and API ecosystems enable users to be reached within days or even hours. This acceleration collapses the temporal buffer that once separated prototyping from production. What used to be a deliberate decision to harden a prototype can be bypassed as systems move from generation to deployment in a single continuous step.

\begin{figure}[ht]
\centering
\includegraphics[width=\linewidth]{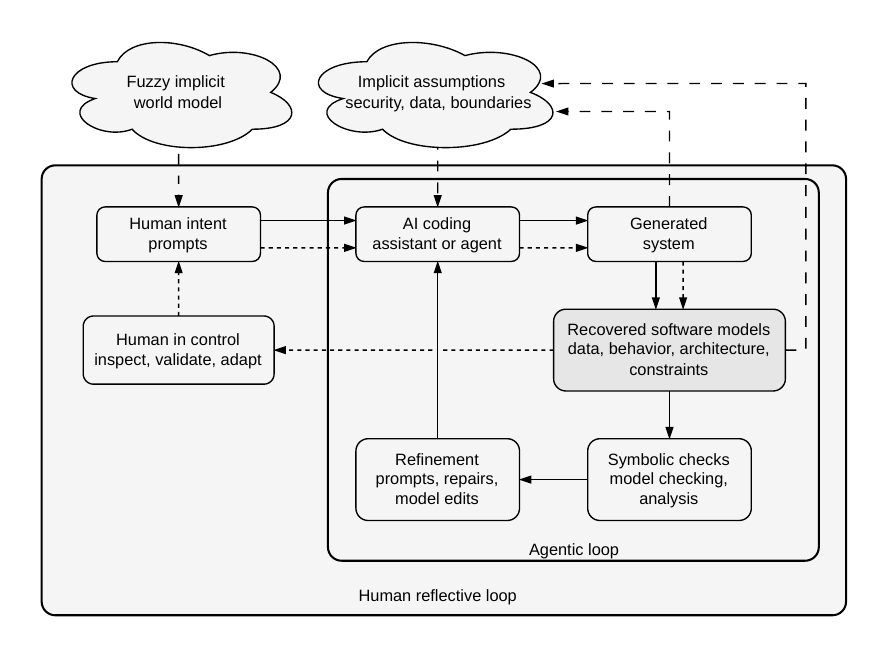}
\caption{Software models as mediators in an agentic loop between prompt-driven generation, automated checking, refinement, and human reflection.}
\Description{A workflow diagram showing a fuzzy implicit world model feeding human intent and prompts. Human intent flows into an agentic model loop containing an AI coding assistant or agent, a generated system, recovered software models, symbolic checks, and refinement prompts, repairs, and model edits. Recovered models also make implicit assumptions explicit and support human inspection, validation, and adaptation of intent.}
\label{fig:model-mediated-loop}
\end{figure}

\paragraph{\textbf{A Structural Predicament}}
Seen in this light, vibe coding is not simply a new name for prototyping. It represents a structural transformation of how software is conceived, produced, and deployed. The traditional cycle of requirements, design, implementation, and testing assumes that models and abstractions exist before code. Vibe coding inverts this order. Code comes first, models follow later if they exist at all. The result is a new epistemic landscape in which understanding lags behind generation. Developers, users, and even the creators themselves struggle to make sense of what has been produced~\cite{PratherR0MRBKWB24}.

This structural inversion creates the conditions for the kinds of failure described earlier. Without explicit abstractions to mediate between intention and implementation, the system remains opaque~\cite{bencomo2024abstractionengineering}. The problem is not simply that vibe coding is bad prototyping. It is the act of generation without understanding through modeling that displaces the foundation upon which robust, maintainable engineering practices have traditionally rested~\cite{eckert2022models, dickerson2013brief}.

\section{Software Models as Mediators}

\paragraph{\textbf{From Prequels to Post-Hoc Artifacts}}
Conceptual models have traditionally been employed as prequels to programming. They served as blueprints, abstractions that guided implementation and mediated communication among stakeholders~\cite{mylopoulos1992conceptual}. These models did not merely describe software. They prescribed how it should be built. Their epistemic role was anticipatory. They structured thinking and formalized communication in advance of coding~\cite{GuizzardiPS23}.

As argued above, generative AI often inverts this sequencing. Functionality is expressed in natural language, generated as code, and only later subjected to engineering scrutiny. The modeling question therefore changes. Instead of asking only how models can guide implementation, we must also ask how models can recover and stabilize understanding after implementation.

\paragraph{\textbf{Model-Mediated Feedback}}
We propose model-mediated feedback loops for AI-generated software. They treat generated code as an intermediate step in engineering, not as the final artifact. Fig.~\ref{fig:model-mediated-loop} shows how these operations support an \textit{automatic agentic loop} and a \textit{human reflective loop}.

\paragraph{\textbf{Two Complementary Loops}}

In the \textit{agentic loop}, generated code is not merely handed back to an AI assistant for another edit. It is abstracted into expected models of data, behavior, architecture, runtime context, or constraints. These models provide a symbolic handle on the generated system. Model checking, program analysis, tests, or policy checks can then compare the expected model with the implementation and produce concrete feedback. That feedback can become a repair prompt, a generated regression test, a static check, a model edit, or an instruction that constrains later agent actions. Recovered model elements remain linked to their supporting code, tests, or traces and distinguish observed behavior from inferred or asserted constraints. Once those constraints are validated by a human or authoritative policy, violations can produce evidence-linked counterexamples, regression tests, or repair prompts; recovery and checking then repeat after repair.

The \textit{human reflective loop} uses the same recovered models in different ways. Here, the models are not primarily inputs to an automatic repair step. They represent assumptions that were implicit in the human's world model, the prompt, the agent's interpretation, and the generated code. By inspecting the recovered model, the human can notice that a concept is missing, a boundary is wrong, or a security assumption was never made explicit. The resulting feedback goes back into the human's intent and into the next prompt or model edit. This makes the model a mediating artifact between the human in the control seat and the agentic generation loop.

\begin{table*}[t]
\caption{Candidate software models for post-hoc mediation in AI-generated systems.}
\Description{A table listing model types, recovery sources, uses, and risks surfaced.}
\label{tab:model-types}
\small
\begin{tabular}{@{}p{0.16\textwidth}p{0.24\textwidth}p{0.29\textwidth}p{0.22\textwidth}@{}}
\toprule
\textbf{Model type} & \textbf{Recovery sources} & \textbf{Primary use} & \textbf{Example risk surfaced} \\
\midrule
Domain and data model & Database schemas, ORM classes, API payloads, logs & Help creators, developers, and agents inspect core concepts, relationships, and business invariants & Missing relation between subscription state, usage records, and payment status \\
Behavioral model & Routes, traces, tests, event logs, controller code & Check request ordering, protocols, and state changes & External API call occurs before server-side authorization \\
Architectural model & Imports, dependencies, service boundaries, deployment manifests & Reveal component boundaries and isolate responsibilities & Secrets or database access leak into client-facing code \\
Constraint model & Assertions, policies, tests, prompts, inferred invariants & Turn intent into checks, repair prompts, or generation constraints & Paid access, rate limits, or data ownership rules are implicit \\
Runtime model & Infrastructure configuration, environment variables, monitoring data & Connect generated code to operational assumptions & Missing rate limits, exposed keys, or unsafe deployment defaults \\
\bottomrule
\end{tabular}
\end{table*}

\paragraph{\textbf{Model Types and Workflow Fit}}
Table~\ref{tab:model-types} makes the loop more concrete. The goal is not to recover every possible model or to impose heavyweight modeling before a user can continue working. Instead, the loop should appear at natural pauses in a vibe coding workflow. It can run before deployment, after an agentic change, when tests fail, when a prototype gains users, or during review. At these moments, recovered models provide a compact representation that is easier to inspect than a large prompting history and the correspondingly generated codebase. The value of these models is strongest when they remain linked to code, tests, traces, and prompts, so that users can move from an abstract concern to the artifact that supports or violates it. This does not make model recovery trivial. Reverse engineering design intent from code remains incomplete and uncertain, especially when generated code contains ad hoc structure or when the rationale behind prior changes is unavailable~\cite{burge2003rationale,liu2025code,wang2025beyond}. The challenge is therefore to recover useful partial models with explicit evidence and uncertainty rather than to promise perfect reconstruction. For humans working under time pressure or with verbose AI-generated feedback, such models should reduce the amount of code and prompting history that must be inspected before a judgment can be made~\cite{kuutila2020time,alami2025human}.

\paragraph{\textbf{A Small Hardening Scenario}}
Consider a generated SaaS application with users, subscriptions, API keys, payments, usage records, and an external AI service. The application works in a demonstration. A user signs up, enters payment details, receives access, and sends requests to the external service. Later, attackers discover that subscription checks can be bypassed and that usage can be charged to exposed keys.

Although Table~\ref{tab:model-types} presents a broader repertoire, a first realization can focus on the behavioral and constraint models. The recovered behavioral model records the request path from authentication and subscription checking to usage recording and the external API call. Each transition links to the route or controller code, test, or trace that supports it. It can thereby show that the external call occurs before authorization, that usage is not associated with the authenticated user, or that a check exists only in client-side code. These are only \emph{observations} about the implementation. In this concrete case, whether paid access and ownership are required remains an inferred assumption until the creator or an authoritative policy confirms it.

The constraint model then turns these observations into an explicit candidate rule. Every external API call must be preceded by a server-side subscription check and must be associated with a usage record owned by the authenticated user. In the human reflective loop, the rule gives the creator a reviewable statement of what paid access means and where that assumption must hold. At this point, it remains a hypothesis about intent rather than a guarantee. Once the creator or an authoritative policy validates it, the rule can be encoded in tests or checks and enforced within their stated coverage. In the agentic loop, the validated rule is checked against the behavioral model. A violation produces a counterexample linked to the offending transition and source evidence; this can generate a regression test and a repair prompt that moves the check to the server route or corrects usage ownership. Recovery and checking then run again on the repaired code. The model is thus an artifact through which intent is proposed, reviewed, validated, and operationalized.

\paragraph{\textbf{Models as Coordination Artifacts}}
The same mechanism also explains why models matter for agentic workflows~\cite{hassan2025AgenticSE}. Agents that monitor, test, repair, or optimize software need shared context. Raw code is too fine-grained and unstable to serve as the only coordination medium~\cite{wang2025interaction}. Models can provide structured, machine-readable representations that complement LLMs' local code synthesis with global structure~\cite{Gu-ChallengesAIAdoptionInSe}. Domain models anchor vocabulary, behavioral models constrain interactions, architectural models expose boundaries, and constraint models make invariants explicit. In this role, models are useful to humans and to AI systems alike.

\paragraph{\textbf{Toward a New Epistemic Role}}
Reimagined in this way, models are no longer only prequels to code. They become mediators of intent, comprehension, refinement, and evolution. The resulting lifecycle does not return us to traditional model-driven engineering. It opens a trajectory in which models remain central, but their centrality comes from making generated systems understandable and governable after code already exists.

\section{Future Plans}
The proposed role of models raises several challenges for software engineering and modeling research. We frame them as future plans for software models rather than as an agenda only for model-driven engineering, because the problem sits between current AI-assisted programming practice and the long tradition of model-based reasoning. The central question is how models can become useful working artifacts in AI-driven development without requiring a return to heavy upfront modeling.

\paragraph{\textbf{Fidelity and Traceability}}
Recovered models must be faithful enough to support decisions without implying more order than the generated system contains. AI-generated code is often inconsistent, redundant, or ad hoc~\cite{liu2025code, wang2025beyond}. A recovered model may simplify this code in useful ways, but it can also hide defects by making an accidental structure look intentional. A concrete research task is to attach evidence to model elements, such as source locations, tests, logs, traces, prompts, and confidence markers. Fidelity measures can then distinguish observed facts from inferred intent and unsupported assumptions~\cite{MunozWTV24}. This limit is structural, not merely an engineering problem. Evidence that is absent in code and traces cannot be reconstructed reliably, and a recovered model cannot provide assurance equivalent to a validated forward model. This is especially consequential in regulated, safety-critical, or certification settings, where original requirements, rationale, and provenance form part of the assurance case. Recovered models can support comprehension and heuristic defect discovery in the middle ground scoped by this paper, but they are not substitutes for upfront specifications and traceable assurance.

\paragraph{\textbf{Model Choice and Abstraction Level}}
Different users need different models. A non-expert creator may need a domain model and a few critical invariants. A developer may need architectural and behavioral views. An AI repair agent may need machine-readable constraints and traceability to code. The next step is to compare model types such as class, component, sequence, state, data-flow, deployment, and constraint views for concrete decisions in the feedback loop. Some cases can reuse established UML or architectural notations. Others may need lighter, formalized, partial, or conversational forms of modeling that expose only the decisions at stake~\cite{FillC0S24,JongelingCCC22}.

\paragraph{\textbf{Verification and Repair}}
Models can surface risks that are hard to see in raw code, including missing authorization paths, inconsistent schema use, architectural erosion, and model smells~\cite{MumtazAMN19,MisbhauddinA15,SmajevicB21,SmajevicHB21}. This promise creates a responsibility to verify the models themselves. A wrong model can produce misplaced confidence. Verification should therefore check recovered models against code, tests, runtime traces, documentation, blueprints, and developer intent where available. Repair support should then translate model-level findings into concrete artifacts, including regression tests, static checks, repair prompts, pull request comments, or model transformations that guide later generation.

\paragraph{\textbf{Usability for Non-Experts}}
The democratizing appeal of vibe coding should not be lost by reintroducing models in a form that only experts can use. Prior work shows that practitioners often use UML selectively and informally rather than as a complete formal language~\cite{Petre13,JongelingC24,LaraG25}. Model-mediated AI development should build on this reality~\cite{MODELS25.NIER}. Useful interfaces may combine lightweight diagrams, natural-language explanations linked to model elements, progressive disclosure of detail, and navigation from a model-level concern to the code or test that implements it. The research problem is not only notation design, but also when and how to present a model so it supports judgment without interrupting creative flow.

\paragraph{\textbf{Process Integration}}
Models will matter only if they fit into the workflows where AI-generated software is produced. They should appear at moments when they help users make decisions, such as before deployment, after agentic code changes, during security review, or when a prototype gains real users. This requires tool support that connects modeling environments, IDEs, AI coding assistants, test systems, and repositories~\cite{Russo24}. One practical direction is to treat model recovery as a CI or repository service that produces a compact review artifact whenever an agent changes code or whenever generated software crosses a deployment threshold. Recent advances in web-based modeling tool development~\cite{BorkLO23,GuerraL18,MODELS26-MCP4GLSP} should facilitate the integration of model generation, rendering, and analysis into the AI-driven software engineering workflow.

\paragraph{\textbf{Evaluation}}
This proposal needs empirical evaluation beyond plausible examples. Studies should ask whether recovered models help users identify missing authorization, invalid data ownership, architectural boundary violations, or unsafe deployment assumptions faster and more accurately than code review alone. They should also test whether models improve comprehension of generated systems, support maintainability and evolution tasks, reduce unnecessary inspection effort, and increase justified confidence in AI-assisted changes. Evaluations should compare expert developers, junior developers, and non-expert creators, because the useful level of abstraction will likely differ across groups. Such studies would turn the proposal from a conceptual reframing into measurable evidence about when software models improve AI-assisted engineering.

\section{Conclusion}

Vibe coding and agentic software development illustrate both the promise and the risk of generative AI in software engineering. They make software creation more accessible, but they also allow prototypes to become consequential systems before their structure, assumptions, and risks are fully understood. We have argued that software models can help address this gap by taking on a renewed mediating role. In the agentic loop, recovered and refined models give AI systems symbolic feedback for checking, repair, and future constraints. In the human reflective loop, the same models give creators and developers inspectable representations of assumptions that would otherwise remain buried in prompts and generated code.

This is a bridge between software engineering practice and model-based reasoning. It does not require a full return to upfront modeling, nor does it treat models as static documentation. It treats them as working artifacts that connect intent, code, validation, and evolution. If vibe coding and agentic development are becoming a new form of prototyping, software models may be the artifacts that help turn generated prototypes into systems that can be understood, hardened, and maintained.

\begin{acks}
This research was partially funded by the Austrian Science Fund (FWF) under grant
10.55776/PIN3275223 (\url{https://doi.org/10.55776/PIN3275223}) and by the Austrian Research Promotion Agency (FFG) through the Smart GLSP project (grant FO999925707).
\end{acks}

\bibliographystyle{ACM-Reference-Format}
\bibliography{references}

\end{document}